\documentclass[sigplan,screen]{acmart}

\usepackage[frozencache,cachedir=.]{minted}
\usepackage{graphicx}
\usepackage[ruled,linesnumbered]{algorithm2e}
\usepackage{subcaption}
\usepackage{multirow}
\usepackage{rotating}
\usepackage{array}  
\usepackage{colortbl}
\usepackage{fontawesome}
\usepackage{hyperref} 
\usepackage{tikz}
\usepackage{lipsum}
\usepackage{xfrac}
\usepackage{xcolor}
\usepackage{color}
\usepackage{mdframed}
\usepackage[export]{adjustbox} 
\usepackage{float}  
\usepackage{scalerel}
\usepackage{tcolorbox}
\usepackage{marvosym}

\setminted[python]{
  fontsize=\fontsize{9pt}{11pt}\selectfont,
  fontfamily=lmtt,
}

\newcommand{\ExternalLink}{%
    \tikz[x=1.2ex, y=1.2ex, baseline=-0.05ex]{%
        \begin{scope}[x=1ex, y=1ex]
            \clip (-0.1,-0.1) 
                --++ (-0, 1.2) 
                --++ (0.6, 0) 
                --++ (0, -0.6) 
                --++ (0.6, 0) 
                --++ (0, -1);
            \path[draw, 
                line width = 0.7, 
                rounded corners=0.5] 
                (0,0) rectangle (1,1);
        \end{scope}
        \path[draw, line width = 0.7] (0.5, 0.5) 
            -- (1, 1);
        \path[draw, line width = 0.7] (0.6, 1) 
            -- (1, 1) -- (1, 0.6);
    }
}

\newcommand\marklessfootnote[1]{
    \addtocounter{footnote}{1} 
    \footnotetext{#1}
}

\copyrightyear{2025}
\acmYear{2025}
\setcopyright{acmlicensed}
\acmConference[ASPLOS '25]{Proceedings of the 30th ACM International Conference on Architectural Support for Programming Languages and Operating Systems, Volume 2}{March 30-April 3, 2025}{Rotterdam, Netherlands}
\acmBooktitle{Proceedings of the 30th ACM International Conference on Architectural Support for Programming Languages and Operating Systems, Volume 2 (ASPLOS '25), March 30-April 3, 2025, Rotterdam, Netherlands}
\acmDOI{10.1145/3676641.3716008}
\acmISBN{979-8-4007-1079-7/2025/03}

\begin{document}

\title[Orion: A Fully Homomorphic Encryption Framework for Deep Learning]{Orion: A Fully Homomorphic Encryption \\ Framework for Deep Learning}

\author{Austin Ebel$^*$}
\affiliation{
  \institution{New York University}
  \city{Brooklyn}
  \state{NY}
  \country{USA}
}
\email{abe5240@nyu.edu}

\author{Karthik Garimella$^*$}
\affiliation{
  \institution{New York University}
   \city{Brooklyn}
  \state{NY}
  \country{USA}
}
  \email{kg2383@nyu.edu}

\author{Brandon Reagen}
\affiliation{
  \institution{New York University}
  \city{Brooklyn}
  \state{NY}
  \country{USA} 
}
\email{bjr5@nyu.edu}

\pagenumbering{arabic}
\begin{abstract}
\label{abs}

Fully Homomorphic Encryption (FHE) has the potential to substantially improve privacy and security by enabling computation directly on encrypted data. This is especially true with deep learning, as today, many popular user services are powered by neural networks in the cloud. Beyond its well-known high computational costs, one of the major challenges facing wide-scale deployment of FHE-secured neural inference is effectively mapping these networks to FHE primitives. FHE poses many programming challenges including packing large vectors, managing accumulated noise, and translating arbitrary and general-purpose programs to the limited instruction set provided by FHE. These challenges make building large FHE neural networks intractable using the tools available today. 

In this paper we address these challenges with \textit{Orion}, a fully-automated framework for private neural inference using FHE.
Orion accepts deep neural networks written in PyTorch and translates them into efficient FHE programs. We achieve this by proposing a novel single-shot
multiplexed packing strategy for arbitrary convolutions and through a new, efficient technique to automate bootstrap placement and scale management. We evaluate Orion on common benchmarks used by the FHE deep learning community and outperform state-of-the-art by $2.38 \times$ on ResNet-$20$, the largest network they report.
Orion's techniques enable processing much deeper and larger networks. We demonstrate this by evaluating ResNet-50 on ImageNet and present the first high-resolution FHE object detection experiments using a YOLO-v1 model with $139$ million parameters. Orion is open-source for all to use at: \href{https://github.com/baahl-nyu/orion}{\fontfamily{lmtt}\selectfont https://github.com/baahl-nyu/orion}.

\end{abstract}

\renewcommand{\shortauthors}{Austin Ebel, Karthik Garimella, \& Brandon Reagen}



\maketitle 

\marklessfootnote{Equal contribution. \scalebox{1.25}{\Letter} Austin Ebel is the corresponding author.}

\vspace{-1.9px}

\section{Introduction}
\label{sec:intro}

Fully Homomorphic Encryption (FHE) is a powerful encryption scheme that allows for computation to be performed directly on encrypted data without ever needing to decrypt. In this manner, both the original data and any intermediate results remain encrypted and can only be observed in the clear when the data owner decrypts it locally using their secret key. FHE provides a solution to the security and privacy concerns of outsourced cloud computation. This has broad implications for areas such as finance or health and other situations involving sensitive data. Many of these services are now driven by deep learning, specifically neural networks. 

Today, the wide-scale deployment of FHE-enabled private deep learning remains limited due to the significant computational costs of FHE and the programming challenges of translating standard, unencrypted neural networks into FHE programs. While hardware accelerators have addressed this performance gap~\cite{reagen, feldmann2021f1, Kim_2022, CraterLake, soni2023rpu, sharp, osiris} and FHE compilers have made progress in translating workloads into FHE programs ~\cite{porcupine, coyote, CHET, EVA, viand2023heco, fhelipe}, it remains a challenge to run FHE neural inference on large datasets in a high-level deep learning framework (e.g., ImageNet inference in PyTorch). This paper focuses on efficiently and \textit{automatically} mapping neural networks, as implemented in modern deep learning libraries, to FHE programs. Besides the computational costs, multiple compounding factors make effectively mapping neural networks to FHE challenging. 

\vspace{2px}

\noindent\textbf{Vector packing:} The CKKS FHE scheme \cite{ckks, cryptoeprint:2018/931} encrypts large vectors that typically vary in length from $2^{14}$ to $2^{17}$, with recent work favoring the longer lengths~\cite{soni2023rpu, Kim_2022, dacapo}. Scalar data must be packed into vector \textit{slots}, as CKKS supports only SIMD addition, SIMD multiplication, and cyclic rotation on these encrypted vectors. As a result, performance is directly related to how densely one can fill these slots with meaningful data. Individual encrypted slots cannot be efficiently indexed and any function applied to a vector is applied to all its elements. Therefore, it is crucial to optimize data packing and layout in CKKS vectors to efficiently make use of the available slots and leverage this SIMD property. 

\vspace{2px}
\noindent\textbf{Bootstrap placement:} Each encrypted vector can only execute a finite number of computations before decryption fails. To address this shortcoming, bootstrapping can be used to reset the computational budget but comes at an extremely high latency cost~\cite{DHBSGS}. Effectively inserting bootstrap operations into an FHE program is challenging; bootstrapping at any point in an FHE circuit directly affects the latency of subsequent operations as well as the placement of future bootstraps. Therefore, an efficient solution to bootstrap placement requires a holistic understanding of the FHE program.

\vspace{2px}
\noindent\textbf{Programmability:} Existing FHE libraries are not designed with machine learning in mind and are cumbersome for building large FHE programs. Every program must be translated into a series of SIMD additions, SIMD multiplications, and cyclic rotations on very large vectors. This translation is especially challenging for deep learning since encrypted convolutional and fully-connected layers cannot be evaluated using traditional tensor arithmetic, and because element-wise nonlinear activation functions must be approximated with high-degree polynomials.
 Finally, several auxiliary FHE operations (e.g., rescaling, bootstrapping, and level adjustment) must be manually inserted into FHE programs making it hard for 
 practitioners to immediately make use of FHE. 

To address these challenges, we develop and open-source \textbf{Orion}: a fully-automated framework for private neural inference that advances the state-of-the-art in all three areas. For data packing, we introduce a novel packing strategy for convolutions that we call \textit{single-shot multiplexing}. This approach directly improves upon the multiplexed packing technique of Lee et al. \cite{lee2022} in three ways: (i) it halves the multiplicative depth of every strided convolution from two to one, (ii) it supports convolutions with arbitrary parameters (e.g., stride, padding, groups, and dilation), and (iii) it reduces expensive ciphertext rotations in modern 
networks by up to $6.41\times$. 

We then present a fully-automated and efficient solution for placing bootstrap operations within a neural network, which we express as a directed acylic graph (DAG). Our approach finds the shortest paths (i.e., latencies) through related DAGs to determine the locations of bootstrap operations that abide by the constraints of FHE.
Our algorithm scales linearly with network depth and matches state-of-the-art \cite{dacapo} in terms of number of bootstraps, while also being between $8\times$ and $1270\times$ faster. As an example, it takes Orion just $1.94$ ($11.0$) seconds to place bootstraps in ResNet-20 (ResNet-110) and requires no user input. Notably, our method also places just $351$ bootstrap operations in a ResNet-50 evaluated on ImageNet, whereas $8480$ bootstraps are placed in HeLayers \cite{helayers2}. For ResNet-20, Fhelipe \cite{fhelipe} (PLDI '24) performs $58$ bootstraps, whereas we have only $37$. 

Finally, we design Orion to be accessible to machine learning researchers and practitioners who are unfamiliar with FHE. Orion inherits directly from PyTorch and extends the functionality of its most popular CNN layers. For instance, users can train Orion networks with existing PyTorch training scripts and directly load the weights of pre-trained models with \texttt{torchvision}. Orion's interoperability with PyTorch allows us to easily verify that our FHE outputs match the outputs of running inference directly in PyTorch. We demonstrate our capabilities by running, to the best of our knowledge, the largest FHE inference to date: YOLO-v1 \cite{yolov1} on the PASCAL-VOC dataset \cite{pascalvoc}.

We evaluate Orion using a set of neural networks and datasets and report significant improvements over state-of-the-art \cite{lee2022, dacapo, fhelipe}. 
We make the following contributions:

\begin{itemize}
    \item[1.] We develop and open-source Orion: a fully-automated framework for running FHE neural network inference directly in PyTorch. Using Orion on the standard ResNet-20 FHE benchmark, we achieve a speedup over state-of-the-art (Fhelipe~\cite{fhelipe}) by $2.38 \times$. 
    \item[2.] We present the \textit{single-shot multiplexed} packing strategy in Orion and support arbitrary convolutions while only having a multiplicative depth of one.
    \item[3.] We implement our automatic bootstrap placement algorithm in Orion, which requires no user input, and matches the state-of-the-art \cite{dacapo} in terms of number of bootstraps while running up to $1270\times$ faster.
    \item[4.] We are the first to perform high-resolution object detection using a YOLO-v1 model with $139$ million parameters on an image of size  $448 \times 448 \times 3$ \cite{yolov1}; to this best of our knowledge, this is the largest FHE computation to date.
\end{itemize}
\section{CKKS Background}
\label{sect:background}

In this section, we describe the CKKS homomorphic encryption scheme  used in Orion. At a high level, CKKS encrypts large vectors of complex (or real) numbers and supports three operations: element-wise addition, element-wise multiplication, and cyclic rotation \cite{ckks}. These properties make CKKS a natural choice for applications that operate on real-valued vectors such as deep learning. Below, we detail the fundamental datatypes and operations in CKKS. 
Please refer to Table \ref{tab:ckks} for our notation.

\vspace{-5px}

\subsection{Datatypes}
\label{subsect:dtypes}
The three datatypes in CKKS are \textit{cleartexts}, \textit{plaintexts}, and \textit{ciphertexts}. A cleartext is an unencrypted vector of complex (or real) numbers. Encoding converts a cleartext into a plaintext, which is still unencrypted and is an element of the polynomial ring $\mathcal{R}_Q = \mathbb{Z}_Q[X]/(X^N + 1)$. Here, $N$ is a power of two and each coefficient is an integer modulo $Q$. For technical reasons, the modulus $Q$ is typically chosen to be a product of $L+1$ primes $q_i$ such that $q_i \bmod 2N = 1$ and $\log_2 q_i \approx w$ where $w$ is a standard machine-word size (32 or 64) \cite{cryptoeprint:2018/931}.  In this way, a plaintext is a polynomial up to degree $N-1$ and each coefficient is in $\mathbb{Z}_Q$. 

\setlength{\tabcolsep}{4pt}
\renewcommand{\arraystretch}{1.2}
\begin{table}[t]
    \small
   \caption{Relevant CKKS parameters.}
   \vspace{-1em}
   \label{tab:params}
   \centering
   \begin{tabular}{cl}
   \toprule\toprule
   \textbf{Param.}                            & \textbf{Description} \\
   \midrule
   $ \mathbf{m} $                               & A cleartext vector of real or complex numbers. \\
   $ [ \mathbf{m} ] $                           & A plaintext polynomial encoding $\mathbf{m}$.\\
   $[\![ \mathbf{m} ]\!]$                     & A ciphertext encrypting the plaintext $[ \mathbf{m} ]$.\\
   $N$                                        & Power-of-two polynomial ring degree. \\ 
   $n$                                        & Length of the vector message, $n \leq N$ (or $\sfrac{N}{2}$). \\
   $L$                                        & Maximum multiplicative level of a ciphertext. \\
   $\ell$                                     & Current multiplicative level. \\
   $Q$                                        & Initial polynomial modulus.\\
   $q_i$                                      & Small moduli in RNS decomposition of $Q = \prod_{i=0}^{L} q_i$. \\
   $\Delta$                                   & Scaling factor for $ [ \mathbf{m} ]$ and $[\![ \mathbf{m} ]\!]$.\\
   $L_{\text{boot}}$                          & Number of levels consumed by bootstrapping. \\
   $L_{\text{eff}}$                           & Maximum achievable level after bootstrapping. \\
   \bottomrule
   \end{tabular}
   \label{tab:ckks}
\end{table}

A plaintext can then be encrypted into a ciphertext that consists of two polynomials (i.e., an element in $\mathcal{R}_Q \times \mathcal{R}_Q$). To maintain 128-bit security and enable deep arithmetic computations in CKKS, it is common for $N$ to be between $2^{14}$ to  $2^{17}$ and $Q$ to be on the order of hundreds to thousands of bits ~\cite{cryptoeprint:2019/939}. In practice, these constraints mean plaintexts and ciphertexts are KBs to MBs in size. CKKS performs polynomial operations on both plaintexts and ciphertexts that correspond to SIMD complex addition, SIMD complex multiplication, and cyclic rotations on the underlying cleartext vectors. For our use case, we encrypt real-valued vectors rather than complex-valued vectors and therefore utilize encrypted SIMD real addition and SIMD real multiplication.

\subsection{Encoding: Cleartext $\rightarrow$ Plaintext}
\label{subsubsec:enc}
Encoding converts a cleartext vector $ \mathbf{m}  \in \mathbb{C}^{N/2}$ into a plaintext polynomial $[ \mathbf{m} ] \in \mathcal{R}_Q$ by (i) applying an inverse Fast Fourier Transform (iFFT) on the cleartext values, (ii) multiplying each output by a scaling factor $\Delta$, and (iii) rounding each element to the nearest integer. In this manner, one can \textit{pack} $n = N/2$ complex values into a single plaintext. If the underlying data consists only of real numbers (such as our use case), it is possible to pack $n = N$ real values into a single plaintext. Decoding converts a plaintext polynomial back into a cleartext vector by performing the FFT on the polynomial coefficients and dividing each value by the scaling factor $\Delta$. Other encoding and decoding schemes for CKKS exist, but do not preserve the SIMD property on the underlying cleartext values (see Section IV.A of ~\cite{Kim2023}).

\vspace{-5px}

\subsection{Encryption: Plaintext $\rightarrow$ Ciphertext}
A plaintext $[ \mathbf{m} ] \in \mathcal{R}_Q$ is encrypted into a ciphertext $[\![ \mathbf{m} ]\!] \in \mathcal{R}_Q \times \mathcal{R}_Q$ by adding noise $[ \mathbf{e} ] \in \mathcal{R}_Q$ to the plaintext and encrypting via either the secret or public key. Decryption can only occur via the secret key. Both rounding during encoding and the addition of noise during encryption introduce small errors, and so CKKS 
is an approximate homomorphic encryption scheme. However, these approximations tend to be tolerable for deep learning.

\begin{figure}
    \centering
    \includegraphics[width=\linewidth]{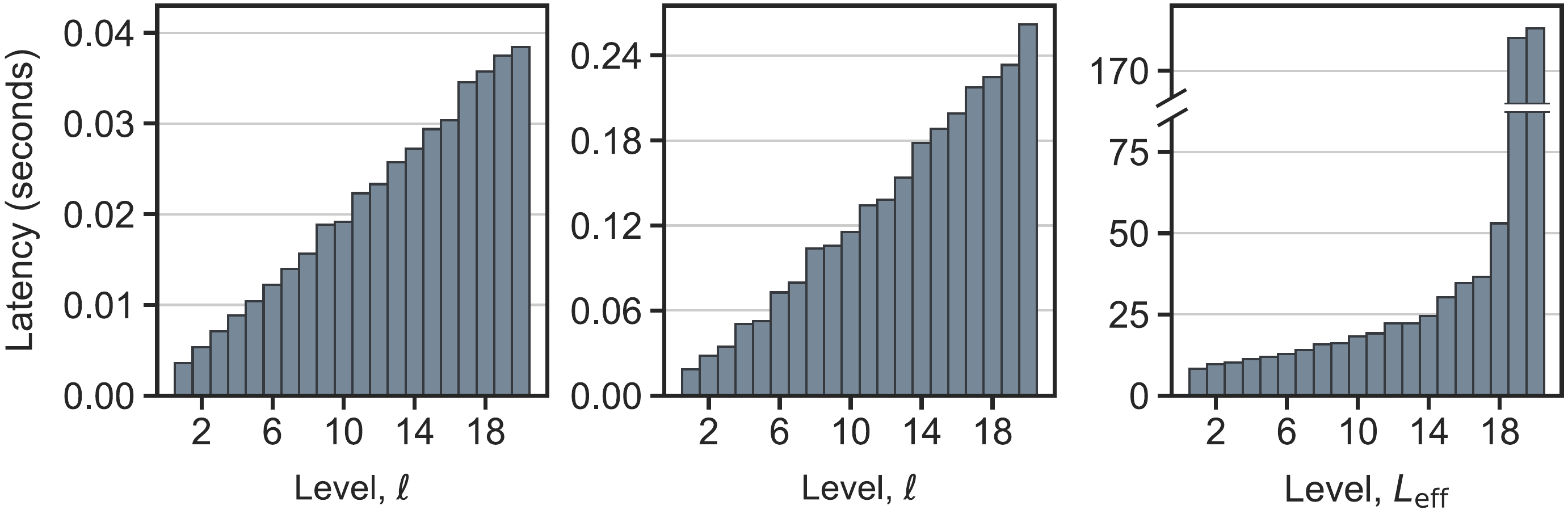}

    \vspace{-5px}
    \hspace{14px}
    \begin{minipage}{0.31\linewidth} 
        \centering
        \subcaption{$\mathsf{PMult}$}
        \label{fig:op_latencies_pmult}
    \end{minipage}
    \hfill
    \begin{minipage}{0.3\linewidth} 
        \centering
        \subcaption{$\mathsf{HRot}$}
        \label{fig:op_latencies_hmult}
    \end{minipage}
    \hfill
    \begin{minipage}{0.3\linewidth} 
        \centering
        \subcaption{$\mathsf{Bootstrap}$}
        \label{fig:op_latencies_boot}
    \end{minipage}

    \vspace{-8px}
    \caption{The latencies of key homomorphic operations as a function of ciphertext level ($\ell$) for ring degree $N = 2^{16}$ and $\Delta \approx 2^{40}$. Setting  $L_{\text{eff}}$ too low would require many low-latency bootstraps, while setting it too high would result in fewer but higher-latency bootstraps. We set $L_{\text{eff}} = 10$ in order to balance bootstrap latency with having a reasonable number of remaining levels for computation.}
    \label{fig:op_latencies}
    \vspace{-10px}
\end{figure}

\vspace{-4px}

\subsection{RNS-CKKS}
\label{subsect:rnsckks}
The coefficients of a CKKS polynomial can be up to thousands of bits wide (e.g., $\log_2 Q \approx 1500$), making operations on these large coefficients compute intensive. Therefore, polynomials are typically decomposed into $L+1$ \textit{residual} polynomials, each with small coefficients that fit within a machine-sized word. In this way, each coefficient in a CKKS polynomial is represented as a vector of length $L+1$, and it follows that the $N$ coefficients for a single CKKS polynomial can be organized as a matrix of size $(L+1) \times N$. Here, each row of the matrix is an $N$-degree residual polynomial with small coefficients (if all $q_i$ are chosen appropriately, then every element of the matrix will fit within a single 32-bit or 64-bit machine word).

This process occurs via the Residual Number System (RNS) in which the large modulus $Q$ is decomposed into $L+1$ smaller moduli $Q = \prod_{i=0}^{L} q_i$. Using RNS, a coefficient $x \in \mathbb{Z}_Q$ can be represented using $L+1$ \textit{limbs} as $x := (x_0, \ldots, x_L)$ where $x_i = x \bmod q_i$. Modular addition (multiplication) between two large elements in $\mathbb{Z}_Q$ is equivalent to performing modular addition (multiplication) on each limb independently ~\cite{Garner1959}.

The integer $L$ is known as the maximum multiplicative level of the polynomial, and as we will see, homomorphic multiplications \textit{consume} levels. At any given time, a ciphertext at level $\ell$ actually has coefficients taken not from $\mathbb{Z}_Q$ but rather from $\mathbb{Z}_{Q_{\ell}}$ where $Q_\ell = \prod_{i=0}^{\ell}q_i$. As a result, the coefficients of a CKKS polynomial at level $\ell$ are actually represented as a matrix of size $(\ell + 1) \times N$ rather than $(L+1) \times N$. For this reason, homomorphic operations at lower levels are less costly. The latency of key homomorphic operations at various levels can be seen in Figure \ref{fig:op_latencies}. When a ciphertext reaches level $\ell = 0$, it has depleted its multiplicative budget, and a \textit{bootstrap} procedure can be used to increase its level up to $L_\text{eff}$ to perform further multiplications. For the interested reader, we note that the superlinear increase in bootstrapping latency observed in Figure \ref{fig:op_latencies}c results from an increase in the decomposition number (\texttt{dnum}) when key-switching to maintain $128$-bit security \cite{hybridks, ciflow}.

\vspace{-5px}

\subsection{CKKS Operations}
\label{subsec:ckks}

We now describe the primitive CKKS operations: addition, multiplication, rotation, and then bootstrapping. 
CKKS performs modular arithmetic on plaintext and ciphertext polynomials to implement these operations. For fast polynomial multiplication, the Number Theoretic Transform (NTT) is used to convert the polynomials from the coefficient to the evaluation representation to reduce the running time from $\mathcal{O}(N^2)$ to $\mathcal{O}(N \log N)$. For this reason, it is common to keep all plaintexts and ciphertexts in the evaluation representation unless otherwise needed. For this section, we will assume we have two cleartext vectors $\mathbf{a}, \mathbf{b} \in \mathbb{R}^n$ that we encode (as plaintexts $[ \mathbf{a} ] $ and $[ \mathbf{b} ] $) both with a scaling factor $\Delta$ and level $\ell$, and further encrypt (as ciphertexts $[\![ \mathbf{a} ]\!]$ and $[\![ \mathbf{b} ]\!]$). 

\vspace{-5px}

\subsubsection{Addition}

CKKS supports point-wise addition between one plaintext and one ciphertext ($\mathsf{PAdd}\left([ \mathbf{a} ], [\![ \mathbf{b} ]\!]\right)$)  or between two ciphertexts (($\mathsf{HAdd}\left([\![ \mathbf{a} ]\!], [\![ \mathbf{b} ]\!]\right)$). Each operand must be at the same number of levels and have the same scaling factor. In both $\mathsf{PAdd}$ and $\mathsf{HAdd}$, the output is a ciphertext $[\![ \mathbf{c} ]\!]_\text{add}$ with the same level and scale as the input operands. This ciphertext corresponds to the SIMD addition of the underlying cleartexts: $\mathsf{decode}(\mathsf{decrypt}([\![ \mathbf{c} ]\!]_\text{add})) \approx \mathbf{a} \oplus \mathbf{b}$. 

\vspace{-5px}

\subsubsection{Multiplication}
\label{subsubsec:mult}
Similar to addition, CKKS supports point-wise multiplication between one plaintext and one ciphertext ($\mathsf{PMult}\left([ \mathbf{a} ], [\![ \mathbf{b} ]\!]\right)$)  or between two ciphertexts ($\mathsf{HMult}\left([\![ \mathbf{a} ]\!], [\![ \mathbf{b} ]\!]\right)$). Each operand must be at the same number of levels, but not necessarily the same scaling factor. Both $\mathsf{PMult}$ and $\mathsf{HMult}$ result in a ciphertext $[\![ \mathbf{c} ]\!]_\text{mult}$ with a scaling factor $\Delta^2$. To prevent exponential growth of the scaling factor, CKKS supports a rescaling procedure that divides the scaling factor by the last prime limb, $q_\ell$, and by choosing each $q_i \approx \Delta$, the scaling factor remains roughly consistent throughout the computation. A key property of the rescaling procedure is that for each coefficient, the component belonging to $\mathbb{Z}_{q_\ell}$ can be (and is) discarded, so that the resulting polynomial has coefficients in $\mathbb{Z}_{Q_{\ell-1}}$ rather than $\mathbb{Z}_{Q_\ell}$; this is why multiplication is said to \textit{consume a level}.

In addition to rescaling, $\mathsf{HMult}$ requires an expensive key-switching operation (that involves auxiliary \textit{evaluation keys}) to ensure correct decryption. Key-switching itself is a computationally expensive procedure involving many NTTs and RNS basis conversions \cite{hybridks, ciflow}. After rescaling (and key-switching for $\mathsf{HMult}$), the output is a ciphertext $[\![ \mathbf{c} ]\!]_\text{mult}$ at level $\ell - 1$ and a scaling factor of $\Delta^2 / q_{\ell} \approx \Delta $. This ciphertext corresponds to the SIMD multiplication of the underlying cleartexts: $\mathsf{decode}(\mathsf{decrypt}([\![ \mathbf{c} ]\!]_\text{mult})) \approx \mathbf{a} \odot \mathbf{b}$.

\vspace{-7px}

\subsubsection{Rotation}
Besides addition and multiplication, CKKS also supports the cyclic rotation ($\mathsf{HRot}_k\left([\![ \mathbf{a} ]\!]\right)$) of the underlying cleartext slots by an amount $0 < k < n$. Similar to $\mathsf{HMult}$, $\mathsf{HRot}$ also requires a key-switching step using \textit{rotation keys} so that decryption of the rotated ciphertext is correct. The resulting ciphertext $[\![ \mathbf{c} ]\!]_\text{rot}$ has the same level and scale as the input ciphertext and corresponds to the underlying cleartext vector cyclically rotated "up" by $k$ slots: $\mathsf{decode}(\mathsf{decrypt}([\![ \mathbf{c} ]\!]_\text{rot}))) \approx (a_k, a_{k+1}, \ldots , a_{k-2}, a_{k-1})$. 

\vspace{-4px}

\subsubsection{Bootstrapping}
\label{ref:bootstrap}
Finally, for a scheme to be fully homomorphic, it must include a way of increasing the number of remaining levels; the bootstrap operation provides this functionality. Bootstrapping is a computationally demanding procedure that increases levels but also consumes a fixed number ($L_{\text{boot}}$) of levels in the process. Therefore, a ciphertext that begins at level $L$ can only reach an effective level, $L_{\text{eff}} = L - L_{\text{boot}}$ after bootstrapping. A typical $L_{\text{boot}}$ is between 13 to 15 levels \cite{DHBSGS, lattigo, HEAAN}.

\section{FHE Matrix-Vector Products}
\label{sect:mv_products}

In this section, we introduce how efficient homomorphic matrix-vector products are performed within Orion.
We focus on the \textit{diagonal method} \cite{halevishoup} of encoding matrices and its state-of-the-art optimizations that include both the baby-step giant-step (BSGS) \cite{halevishoup} and double-hoisting algorithms \cite{DHBSGS}. These are powerful algorithmic optimizations that significantly reduce compute costs. We use these optimizations in \textit{every} linear transformation in Orion including all fully-connected and convolutional layers of neural networks and the matrix-vector products within bootstrapping. Similar to prior work, and in alignment with our threat model, we assume that the weight matrix is a cleartext that can be pre-processed before being encoded, and that the input vector is a ciphertext. 

\vspace{-7px}

\subsection{Diagonal Encoding Method}

\begin{figure}

\subfloat[The diagonal method \cite{halevishoup}. $n=6$ ciphertext rotations are required, one per non-zero diagonal (including the trivial rotation by $0$).]{%
    \label{fig:diagonal}
  \includegraphics[clip,width=\columnwidth]{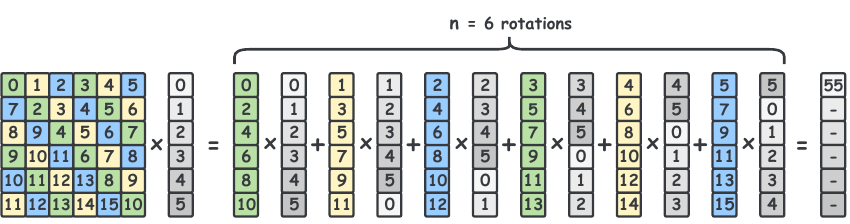}%
}

\subfloat[Extending the diagonal method with BSGS \cite{halevishoup}. Note that only $n_1 + n_2 = 5$ rotations are required and that $n_1 n_2 = n = 6$.]{%
    \label{fig:bsgs}
  \includegraphics[clip,width=\columnwidth]{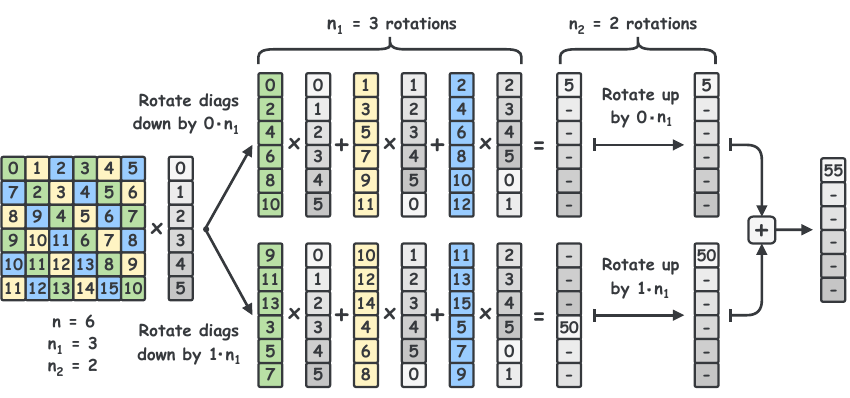}%
}

\vspace{-5px}
\caption{Visualizing how the BSGS algorithm \cite{halevishoup} reduces the number of ciphertext rotations in matrix-vector products.}

\label{fig:mv-prods}
\vspace{-6px}
\end{figure}

The diagonal method \cite{halevishoup} is one technique for performing homomorphic matrix-vector products. This method works by first extracting the \textit{generalized diagonals} of a matrix $\mathbf{M}$, defined as $\mathsf{diag}_{\hspace{0.1em} k} = \mathbf{M}_{[0,\hspace{0.1em}  k]},$ $\mathbf{M}_{[1, \hspace{0.1em} k+1]},$ $\ldots,$ $\mathbf{M}_{[w-1, \hspace{0.1em} k+w-1]}$, where $w$ is the matrix width and the second index is taken modulo $w$. Each diagonal, $\mathsf{diag}_{\hspace{0.1em} k}$, is then multiplied by the input ciphertext, rotated up by $k$ slots, to produce each partial product. Summing all partial products gives the final result. Figure \ref{fig:diagonal} illustrates the diagonal method for a $6 \times 6$ matrix. Notably, this method requires $n$ ciphertext rotations to perform any $n \times n$ matrix-vector product. Thus, its runtime complexity can be seen as $\mathcal{O}(n)$.

\vspace{-5px}

\subsection{Baby-Step Giant-Step Optimization}

The baby-step giant-step (BSGS) algorithm reduces this complexity to $\mathcal{O}(\sqrt{n})$ by leveraging the fact that matrix diagonals can be cheaply rotated before being encoded. More specifically, BSGS decomposes a matrix-vector product into $n_1$ groups of $n_2$ diagonals. Each group shares a common (baby-step) ciphertext rotation, and each diagonal within a group is rotated by a different multiple of $n_1$. The standard diagonal method is then applied across groups to produce $n_2$ partial products, each offset by a multiple of $n_1$. Thus, $n_2$ (giant-step) ciphertext rotations are required to align partial products before they are summed to produce the desired result. Equation (\ref{eq:bsgs}) defines the BSGS algorithm, where $\widetilde{\mathsf{diag}_k} = \mathsf{Rot}_{-\hspace{-0.1em}j \cdot n_1} (\mathsf{diag}_k)$, the $k$'th generalized diagonal of an $n \times n$ matrix, cyclically rotated down by $j\cdot n_1$ slots. Figure \ref{fig:bsgs} also visualizes the BSGS algorithm when $n_1 = 3$, $n_2 = 2$, and $n_1 n_2 = 6$, where diagonals are shaded by group. Notably, one can choose any $n_1$, $n_2$ such that $n_1 n_2 = n$, however the number of ciphertext rotations is minimized when $n_1 = n_2 = \sqrt{n}$.

\vspace{-7px}

{\fontsize{8.99pt}{10pt}\selectfont 
\begin{equation}
    \mathsf{ct}_{\mathsf{out}} = \sum_{j=0}^{n_2 - 1} \mathsf{HRot}_{j \cdot n_1} \left\{ \sum_{i=0}^{n_1 - 1} \mathsf{PMult} \left( \mathsf{HRot}_i (\mathsf{ct}_{\mathsf{in}}), \widetilde{\mathsf{diag}}_{j \cdot n_1 + i} \right) \right\} 
    \label{eq:bsgs}
\end{equation}
}

\subsection{Hoisting Optimizations}

At a high level, \textit{hoisting} is a cryptographic technique that reuses the most expensive aspects of the key-switch procedure over many ciphertext rotations. This optimization is only possible when rotating the \textit{same} ciphertext by different amounts and can therefore only be applied to baby-step rotations in the BSGS algorithm. Bossuat et al. \cite{DHBSGS} separately extend this \textit{single-hoisting} technique to giant-steps and aptly name its implementation \textit{double-hoisting}. In Orion, every matrix-vector product uses the double-hoisting BSGS algorithm (see Algorithm 6 in \cite{DHBSGS}). We refer readers to \cite{DHBSGS,osiris,gazelle} for a more detailed explanation of hoisting.

\vspace{2px}

\noindent\textbf{Takeaway:} The double-hoisting BSGS method has two fundamental benefits: (i) every matrix-vector product consumes just one multiplicative level, and (ii) its performance savings \textit{increase} with increasing matrix size since BSGS decreases the number of ciphertext rotations from $\mathcal{O}(n)$ to $\mathcal{O}(\sqrt{n})$.

\section{Efficient Convolutions in Orion}
\label{sect:packing_convolutions}

In this section, we show how Orion efficiently expresses \textit{all} convolutions as matrix-vector products to leverage the savings from BSGS and hoisting discussed in Section \ref{sect:mv_products}. To achieve this, we introduce a new packing technique called \textit{single-shot multiplexing}. This technique utilizes a modified Toeplitz formulation of convolutions and is highly effective: it supports arbitrary parameters and halves the level consumption of strided convolutions compared to the multiplexed packing approach of the state-of-the-art \cite{lee2022} while also significantly reducing rotation counts. 

Since CKKS operates on vectors of complex (or real) numbers, it is common to first flatten a three-dimensional convolutional input into a row-major (raster-scanned) vector of length $h_i \hspace{0.05em} w_i \hspace{0.05em} c_i$, following the notation from Section \ref{sect:background}. For simplicity, we assume that the number of slots, $n$, matches this length so that all data fits perfectly into a single ciphertext. In Section \ref{sect:single-shot-multiplexing}, we relax this assumption to handle inputs larger than $n$.

\vspace{-5px}

\subsection{SISO Convolutions in Orion}

\begin{figure}
    \centering
    \includegraphics[width=\linewidth]{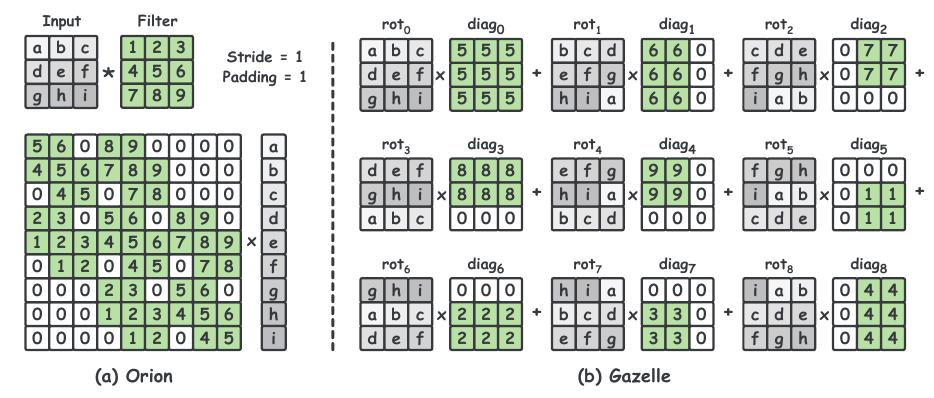}
    \vspace{-20px}
    \caption{An example showing how we transform the packed SISO method from Gazelle (b) into its analogous Toeplitz matrix (a) to leverage the BSGS and hoisting optimizations.}
    \vspace{-8px}
    \label{fig:orion_vs_gazelle}
\end{figure}

We begin with single-input, single-output (SISO) convolutions where  $c_i = c_o = s = 1$  and restrict our attention to same-style convolutions, where the input and output spatial dimensions are equal. An illustrative example is shown in Figure \ref{fig:orion_vs_gazelle}. Gazelle \cite{gazelle} first introduced a packed SISO method to homomorphically evaluate this convolution, and subsequent works \cite{fhelipe, lee2022, dacapo} have continued to follow this blueprint (Figure \ref{fig:orion_vs_gazelle}b).
Here, the input is flattened, packed into a ciphertext, and cyclically rotated $f_h \hspace{0.05em} f_w$ times. Each rotated ciphertext is then multiplied by a \textit{punctured} plaintext, which encodes $n$ copies of a unique filter weight and is selectively zeroed to ensure only the pixels that interact with the filter weight are processed. Finally, the partial products from each multiplication are summed to produce the final output.

\noindent\textbf{Orion:} Our observation here is that prior works' packed SISO method is equivalent to the diagonal method: ciphertexts are rotated and multiplied with pre-processed plaintexts. We show this by working backwards from Figure \ref{fig:orion_vs_gazelle}b to derive its analogous matrix-vector product. In doing so, we arrive at the \textit{Toeplitz} formulation of a convolution, shown in Figure \ref{fig:orion_vs_gazelle}a. In this representation, the filter expands into a weight matrix of size  $h_o \hspace{0.05em} w_o \times h_i \hspace{0.05em} w_i$, where each row corresponds to one filter multiplication. For example, the first row of the matrix in Figure \ref{fig:orion_vs_gazelle}a performs a dot product between the filter weights \texttt{\{5,6,8,9\}} and the input pixels \texttt{\{a,b,d,e\}}, which corresponds to the filter being placed at the top left corner of the input with padding $p=1$. Subsequent rows of the Toeplitz matrix are generated as the kernel slides across the input image. Performing the diagonal encoding method on the matrix in Figure \ref{fig:orion_vs_gazelle}a produces the \textit{exact} operations seen in the packed SISO method in Figure \ref{fig:orion_vs_gazelle}b.

\vspace{2px}

The fundamental benefit of this observation is that we can build a Toeplitz matrix for \textit{any} arbitrary convolution and then apply the double-hoisting BSGS algorithm when evaluating its matrix-vector product. To the best of our knowledge slytHErin \cite{slytherin} is the only prior work that proposes a similar strategy. With this, Orion reduces the number of ciphertext rotations for any SISO convolution from $\mathcal{O}(f)$ to $\mathcal{O}(\sqrt{f})$, where $f$ is the total number of filter elements.

\vspace{-5px}

\subsection{MIMO Convolutions in Orion}

\begin{figure}
    \centering
    \includegraphics[width=\linewidth]{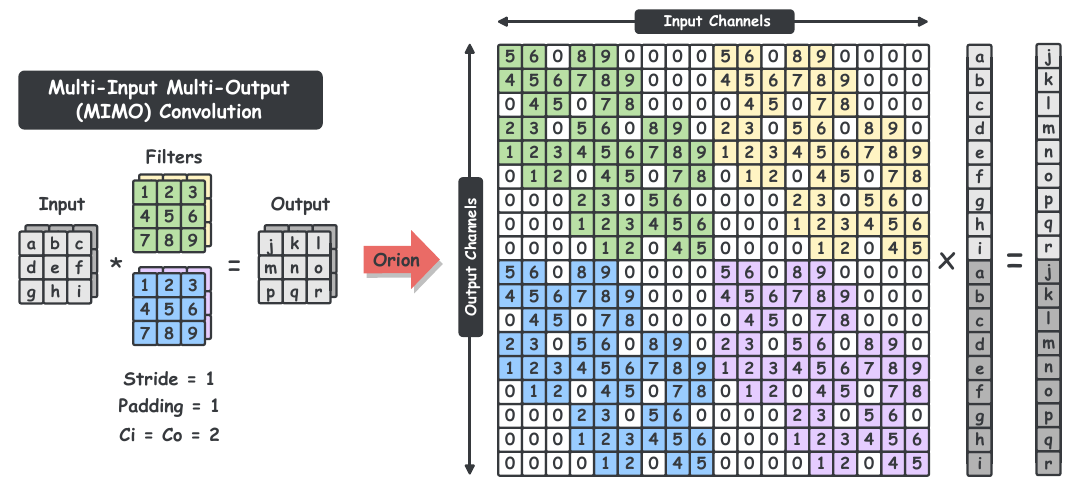}
    \vspace{-15px}
    \caption{The Toeplitz formulation for SISO convolutions from Figure \ref{fig:orion_vs_gazelle} generalizes to MIMO convolutions; filters for separate input (output) channels are placed across the columns (rows) of the Toeplitz matrix.}
    \label{fig:mimo_convolution}
\end{figure}

The Toeplitz formulation extends naturally to both multiple input and multiple output (MIMO) convolutions. Figure \ref{fig:mimo_convolution} shows a MIMO convolution with $c_i = c_o = 2$, along with its analogous Toeplitz matrix. This matrix is constructed in much the same way as the SISO case. Each row represents one filter multiplication, and we can apply the diagonal method to this matrix to produce a raster-scanned output ciphertext in a single multiplicative level.
Once again, this formulation lets us apply double-hoisting BSGS to significantly reduce the number of required ciphertext rotations.

\vspace{-5px}

\subsection{Single-Shot Multiplexed Convolutions}
\label{sect:single-shot-multiplexing}

The Toeplitz approach works well for same-style convolutions because each consecutive row in the matrix shifts the filter by one position over the input. This keeps the same filter element aligned within each matrix diagonal and therefore ensures that the number of expensive ciphertext rotations is \textit{independent} of the input's spatial dimensions. 

\begin{figure*}
    \centering
    \includegraphics[width=0.95\linewidth]{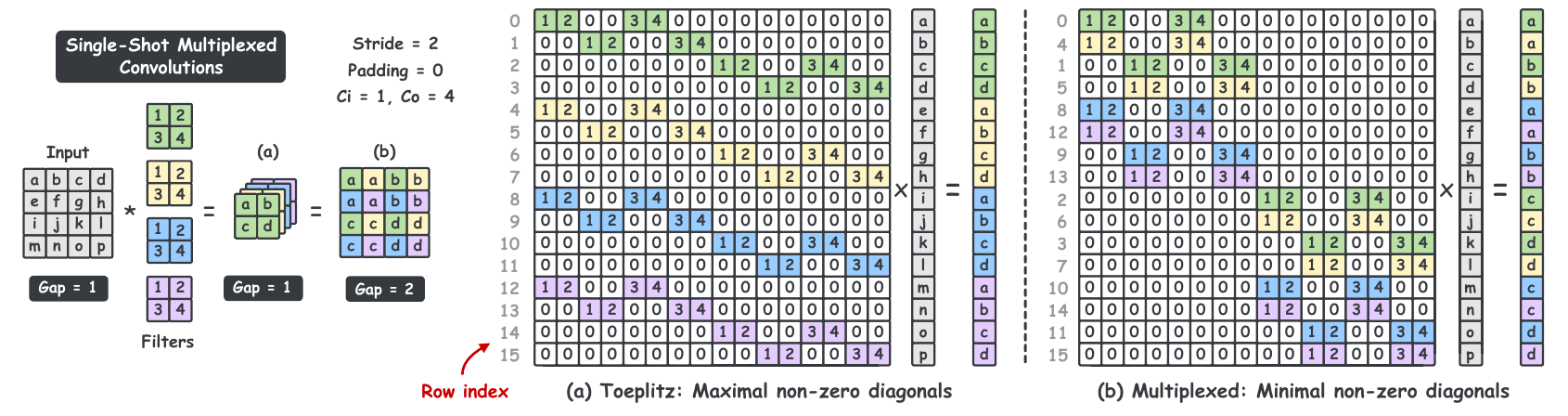}
    \vspace{-10px}
    \caption{Strided convolutions produce Toeplitz matrices with many sparse non-zero diagonals (a). Orion automatically converts these to densely packed multiplexed convolutions to reduce the number of expensive ciphertext rotations (b).}
    \label{fig:strided_convolution}
    \vspace{-5px}
\end{figure*}

However, this convenient property breaks down when considering strided convolutions. Figure \ref{fig:strided_convolution}a illustrates why this occurs using a convolution with a stride of $s=2$. In this case, each successive row in the matrix in Figure \ref{fig:strided_convolution}a shifts the kernel by a multiple of $s$, creating exactly $c_i \hspace{0.05em} h_i \hspace{0.05em} w_i$ non-zero diagonals in the process. Thus, while this method is straightforward, its dependence on the input's spatial dimensions causes it to scale poorly to larger convolutions.

We solve this problem by recognizing that the rows of any Toeplitz matrix can be permuted without changing its underlying computation. This lets us relax the invariant that our ciphertext must represent the image as a flattened tensor. Our approach builds on the multiplexed parallel convolutions introduced by prior work \cite{lee2022}, which interleaves channels within a ciphertext to efficiently handle strided convolutions. Notably, their method consumes two multiplicative levels: one level to first perform a non-strided convolution and a second level to mask and collect only the correct elements (see Figure 5 of \cite{lee2022}). On the other hand, our single-shot multiplexed packing strategy consumes just one level by fusing this mask-and-collect step directly into the weight matrix, which can be pre-processed.

 Figure \ref{fig:strided_convolution}b shows our method for converting the standard Toeplitz matrix in Figure \ref{fig:strided_convolution}a to our single-shot multiplexed solution. Our method produces a densely packed output ciphertext, parameterized by a gap, $g$. 
Subsequent non-strided convolutions maintain this gap, while strided convolutions increase it by a factor of $s$. 
This reformulation lets us leverage both BSGS and double-hoisting and doing so reduces the number of ciphertext rotations in ResNet-20 \cite{he2015deepresiduallearningimage} from $1457$ to $836$. Table \ref{tab:rotation_counts} of our results further compares these two packing strategies in larger neural networks.

\vspace{2px}

\noindent\textbf{Multi-ciphertext:} When an input image does not fit into a single ciphertext (i.e., the number of pixels is greater than the number of slots), the image is split across multiple ciphertexts into contiguous slots where the last ciphertext may only be partially filled with data. In this case, we naturally perform a blocked matrix-vector product with blocks of size $\textit{slots} \times \textit{slots}$. While we evaluate Orion in a single-threaded setting, each block performs independent work and is well-suited for parallel execution across multiple threads or cores.
\vspace{30px}
\section{Automatic Bootstrap Placement}
\label{sect:auto_bootstrap}

In this section, we discuss the problem of bootstrap placement in private neural inference. We begin by highlighting the challenges to motivate the need for automated solutions. Then, we describe the challenges that arise in automation. We close with a general and efficient solution for modern neural networks based on solving shortest paths in directed acyclic graphs (DAGs).

\vspace{-4px}
\subsection{Problem Overview}

Recall from Section \ref{sect:background} that the goal of CKKS bootstrapping is to enable further homomorphic multiplications by increasing a ciphertext’s level from $\ell=0$ to $\ell = L_{\text{eff}} < L$. Determining the optimal location of bootstrap operations within a network is a challenging aspect of FHE inference for three reasons. First, placing a bootstrap at any point in the network directly affects the levels of all subsequent operations, including future bootstrap locations. Second, na\"ive strategies that delay bootstrapping until absolutely necessary often result in more bootstraps being placed, particularly in networks with residual connections (e.g., see Figure 10 of Fhelipe \cite{fhelipe}). Finally, bootstrap runtime grows superlinearly with $L_\text{eff}$, as shown in Figure \ref{fig:op_latencies_boot}. 
Thus, only minimizing the number of bootstrap operations often counter-intuitively increases network latency, since it increases per-bootstrap costs dramatically and  increases the average level (i.e., latency) of all homomorphic operations, as shown in Figures \ref{fig:op_latencies_pmult} and \ref{fig:op_latencies_hmult}. Therefore, a careful balance must be struck to have a high enough $L_\text{eff}$ in order to perform meaningful computation without incurring an exorbitant cost of bootstrapping.

In Orion, our goal is to automatically determine the locations of bootstrap operations to minimize network latency. In doing so, we also determine the levels to perform all linear layers and activation functions. Collectively, we refer to our decisions as a \textit{level management policy}.

\begin{figure*}
    \centering
    \includegraphics[width=0.9\linewidth]{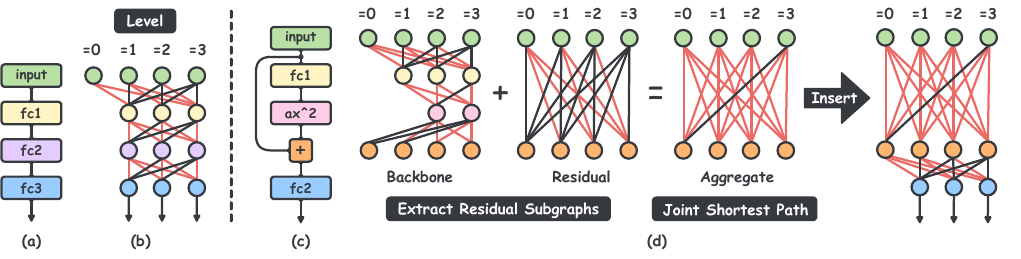}
    \vspace{-8px}
    \caption{A feed-forward neural network with no skip connections (a) and its associated level digraph when $L_\text{eff} = 3$ (b). Edges between nodes connect network layers and red edges correspond to bootstrap operations. This skip-less network does not require a bootstrap operation if the input ciphertext begins at $\ell = 3$. Networks with skip connections (c) require us to first solve the sub-problem along the residual path before holistically solving bootstrap placement for the entire network. This network requires at least one bootstrap.} 
    \label{fig:auto_bootstrap}
    \vspace{-5px}

    \begin{subfigure}{0.45\textwidth}
        \phantomcaption
        \label{fig:level-digraphs-a}
    \end{subfigure}
    \begin{subfigure}{0.45\textwidth}
        \phantomcaption
        \label{fig:level-digraphs-b}
    \end{subfigure}
    \begin{subfigure}{0.45\textwidth}
        \phantomcaption
        \label{fig:level-digraphs-c}
    \end{subfigure}
    \begin{subfigure}{0.45\textwidth}
        \phantomcaption
        \label{fig:level-digraphs-d}
    \end{subfigure}
    \vspace{-10px}
\end{figure*}

\noindent \textbf{Placement constraint:} We restrict the placement of bootstrap operations to occur between network layers, where we define a network layer to be a linear transformation or a polynomial evaluation. Since all linear transforms (e.g., convolutions and fully-connected layers) in Orion consume one level, there is no bootstrap decision to be made within them. On the other hand, activation functions in CKKS must be expressed as a polynomial (e.g., SiLU) or as a composition of several polynomials (e.g., ReLU, which consists of three polynomials \cite{lee2022}) and these approximations tend to consume many levels. We restrict Orion to bootstrap \textit{between} but not \textit{within} polynomial evaluations.

At a high level, a CKKS polynomial evaluation takes as input a single ciphertext and applies the polynomial element-wise to each slot.
The evaluation fans out into a wider circuit of intermediates during the reduction process, which outputs again a single ciphertext. Therefore, if we did choose to bootstrap any of these intermediate ciphertexts, all intermediates at that level would also have to be bootstrapped, increasing bootstrap calls. The reduction steps of polynomial evaluation can be found in prior work \cite{DHBSGS}; we refer the interested reader to Line 4 of Algorithm 1 and Lines $7$, $15$, $17$, $19$ of Algorithm $2$. Thus, we do not bootstrap within polynomials, which also simplies our analysis. Finally, we exclude networks with overlapping skip connections from our analysis, such as DenseNets \cite{densenet}.

\vspace{2px}

\noindent \textbf{Cost model:} Our cost model accounts for the depth of activation layers, the depth and latency of all linear layers, and bootstrap latency. This cost model captures the depth of the circuit (a majority of which comes from the activations) as well as the latency (stemming from the linear layers and bootstrapping). Empirically, we find that $94.1$\% ($95.7$\%) of our ResNet-20 (ResNet-18) latency comes from these two sources.

\vspace{-7px}
\subsection{Orion's Level Management Policy}
\label{subsect:bts_level}

\noindent\textbf{The core idea:} To introduce our level management policy, consider the simple three-layer fully connected network in Figure \ref{fig:level-digraphs-a} without intermediate activation functions. The linear layers $(\texttt{fc1-}\texttt{fc3})$ each consume one level, and we set the effective level ($L_\text{eff}$) to $3$. 
By restricting the placement of bootstrap operations to between network layers, we make it possible to efficiently construct a directed acyclic graph (DAG) that enumerates all possible network states. We refer to these graphs as $\textit{level digraphs}$, and the level digraph for the network in Figure \ref{fig:level-digraphs-a} is shown in Figure \ref{fig:level-digraphs-b}. Each row of nodes in Figure \ref{fig:level-digraphs-b} corresponds to one linear layer. Nodes within a row represent possible choices of level for that linear layer and are weighted by the latency of performing their corresponding linear layer at that level (recall from Figure \ref{fig:op_latencies} that the latencies of homomorphic operations vary as a function of $\ell$). A valid level $\ell$ satisfies $d \leq \ell \leq L_\text{eff}$, where $d$ is the layer’s multiplicative depth. The edges between nodes are weighted by the latency of bootstrapping up to $L_\text{eff}$ if required. In Figure \ref{fig:level-digraphs-b}, we highlight the edges that require bootstrap operations in red.

Note that even when a bootstrap occurs, the subsequent layer can still be performed at $\ell < L_\text{eff}$. While this choice may seem wasteful, the additional degree of freedom often leads to more efficient level management policies, especially in more realistic networks (e.g., ResNet-20). The \textit{optimal} level management policy in Figure \ref{fig:level-digraphs-b} contains the operations along the shortest (lowest total weight) path from any input node ($\texttt{input}_{0 \leq \ell \leq 3}$) to any output node ($\texttt{fc3}_{1 \leq \ell \leq 3}$) in the level digraph, minimizing the latency (with respect to our heuristics) of an inference.

\vspace{2px}
\noindent\textbf{A general strategy:} One problem with the approach in Figure \ref{fig:level-digraphs-b} is that it does not support networks with multiple paths from input to output. Consider the network in Figure \ref{fig:level-digraphs-c} with a residual connection and an activation function. Residual connections are a staple of modern neural network design but complicate the choice of level management policy. A viable solution to Figure \ref{fig:level-digraphs-c} might include solving two independent level digraphs, one per path through the network. However, this approach can lead to clashing choices for the levels of nodes common to both paths. 

Instead, we observe that each residual connection forms a single-entry, single-exit (SESE) region \cite{johnson-sese} in the network graph, bounded by pairs of fork and join nodes. Here, a fork (join) node is any node with more than one child (parent). In Figure \ref{fig:level-digraphs-c}, the nodes $\texttt{\{input\}}$ and $\texttt{\{+\}}$ are the only such fork and join nodes. We leverage this property to extract any SESE region around a residual connection (including its fork and join nodes), treat it as its own distinct problem, and then black-box its solution. To do this, we first construct two separate level digraphs: one for the backbone network and one for the residual connection. Then, we collapse these separate digraphs into one \textit{aggregate} level digraph by solving a joint shortest path problem between \textit{every} pair of input and output nodes. For instance, in Figure \ref{fig:level-digraphs-d}, the weight of the edge in the "aggregate" digraph from $\texttt{\{input\}}_{\ell=0}$ to $\texttt{\{+\}}_{\ell=0}$ is set to the sum of (i) the total weight of the shortest path from  $\texttt{\{input\}}_{\ell=0}$ to $\texttt{\{+\}}_{\ell=0}$ within the "backbone" digraph and (ii) the total weight of the shortest path from $\texttt{\{input\}}_{\ell=0}$ to $\texttt{\{+\}}_{\ell=0}$ within the "residual" digraph. 

Finally, we insert this aggregate level digraph back into the larger network, allowing us to find a shortest path similarly to Figure \ref{fig:level-digraphs-b}. By also tracking the paths summarized by each edge in every aggregate level digraph, we can then work backwards to determine the correct level management policy for all layers, including within the SESE regions that we black-boxed.

\vspace{2px}

\noindent\textbf{Implementation details:} In CKKS, the ReLU activation function is approximated by $x \cdot \mathsf{sign}(x)$, which is itself a SESE region. As ReLUs are often placed within larger residual connections, our approach always begins by black-boxing the inner-most SESE regions before working outwards. 

We estimate the latencies of both the linear layers and bootstrap operations with an analytical model. Then, Orion automatically inserts bootstrap operations within the network, abstracting these complexities away from the end-user. Orion's automatic level management and bootstrap placement algorithm takes only $1.94$ ($0.82$) seconds for ResNet-20 (AlexNet), which is an $8.14\times$ ($1270\times$) speed-up over Dacapo \cite{dacapo} on these same networks. To demonstrate that our solution is tractable, we also ran the algorithm on ResNet-1202, which took just $151$ seconds. Scalability with network depth is further explored in Table \ref{tab:boot_scalability} of our results.
\vspace{-5px}

\setminted[python]{
  fontsize=\fontsize{8.5pt}{9.6pt}\selectfont,
  fontfamily=lmtt,
  escapeinside=||,
}

\definecolor{vibrantblue}{rgb}{0.0, 0.0, 1.0}
\definecolor{black}{rgb}{0.0, 0.0, 0.0}
\definecolor{vibrantred}{rgb}{0.90, 0.40, 0.35}  
\definecolor{matteblack}{rgb}{0.0, 0.0, 0.0}
\definecolor{vibrantpurple}{rgb}{0.68, 0.22, 0.67}  
\definecolor{brightblue}{rgb}{0.19, 0.57, 0.76}  

\renewcommand{\algorithmcfname}{Listing}
{\LinesNumberedHidden
\label{listing:basicblock}
\begin{algorithm}[h]
\caption{ResNet block instantiation in Orion.}
\hspace{-5px}\begin{minipage}{0.90\linewidth}
\begin{minted}[linenos=false]{python}
|\textcolor{vibrantpurple}{import}| |\textcolor{matteblack}{torch.nn}| |\textcolor{vibrantpurple}{as}| |\textcolor{matteblack}{nn}|
|\textcolor{vibrantpurple}{import}| |\textcolor{matteblack}{orion.nn}| |\textcolor{vibrantpurple}{as}| |\textcolor{matteblack}{on}|

|\textcolor{vibrantpurple}{class}| |\textcolor{vibrantblue}{BasicBlock}|(on.Module):
  |\textcolor{vibrantpurple}{def}| |\textcolor{brightblue}{\_\_init\_\_}|(|\textcolor{matteblack}{self}|, Ci, Co, stride=1):
    |\textcolor{brightblue}{super}|().|\textcolor{brightblue}{\_\_init\_\_}|()
    |\textcolor{matteblack}{self}||\textcolor{matteblack}{.conv1 = on.Conv2d(Ci, Co, 3, stride, 1)}|
    |\textcolor{matteblack}{self}|.bn1   = on.BatchNorm2d(Co)
    |\textcolor{matteblack}{self}|.act1  = on.ReLU(degrees=[15,15,27])

    |\textcolor{matteblack}{self}|.conv2 = on.Conv2d(Co, Co, 3, 1, 1)
    |\textcolor{matteblack}{self}|.bn2   = on.BatchNorm2d(Co)
    |\textcolor{matteblack}{self}|.act2  = on.SiLU(degree=63)

    |\textcolor{matteblack}{self}|.add = on.Add()
    |\textcolor{matteblack}{self}|.shortcut = nn.Sequential()
    |\textcolor{vibrantpurple}{if}| stride != 1:
      |\textcolor{matteblack}{self}|.shortcut = nn.Sequential(
        on.Conv2d(Ci, Co, 1, stride, 0),
        on.BatchNorm2d(Co))

  |\textcolor{vibrantpurple}{def}| |\textcolor{brightblue}{forward}|(|\textcolor{matteblack}{self}|, x):
    out = |\textcolor{matteblack}{self}|.act1(|\textcolor{matteblack}{self}|.bn1(|\textcolor{matteblack}{self}|.conv1(x)))
    out = |\textcolor{matteblack}{self}|.bn2(|\textcolor{matteblack}{self}|.conv2(out))
    out = |\textcolor{matteblack}{self}|.add(out, |\textcolor{matteblack}{self}|.shortcut(x))
    |\textcolor{vibrantpurple}{return}| |\textcolor{matteblack}{self}|.act2(x)
\end{minted}
\vspace{-1px}
\end{minipage}
\end{algorithm}
}

\section{The Orion Framework}
\label{sec:orion}

We now present the Orion framework and API, which includes automated support for the optimizations and techniques presented above (e.g., bootstrap placement and packing strategies). 
In addition, Orion automatically handles several subtle, yet difficult, challenges that arise when building large-scale FHE programs: range estimation for high-precision bootstrapping and Chebyshev polynomial evaluation, error-free scaling factor management, and system support for very large data structures.

\vspace{2px}

\noindent\textbf{The Orion API:} FHE neural network applications are notoriously difficult to program, and significant research efforts ~\cite{porcupine, coyote, CHET, EVA, viand2023heco, fhelipe} have focused on reducing this barrier to entry.
Yet, despite this progress, no
high-level, high-performance API exists, which practitioners have come to expect from modern tools like JAX or Pytorch \cite{jax2018github, pytorch}. 
Orion provides this interface.
We implement custom Python bindings to the Lattigo \cite{lattigo} FHE library that directly interact with PyTorch tensors. Then, we use these bindings to build Orion modules that inherit and extend the functionality of their corresponding PyTorch modules. For example, Listing 1 shows a ResNet block in Orion that closely mirrors the equivalent PyTorch implementation (\hspace{0.1em}{\Large{\href{https://github.com/kuangliu/pytorch-cifar/blob/master/models/resnet.py#L14}{\textcolor{black}{\ExternalLink}}\hspace{-0.4em}}} ). In more detail, we make use of our custom $\texttt{orion.nn}$ module (rather than $\texttt{torch.nn}$) when instantiating specific layers. Within $\texttt{orion.nn}$, activation functions accept a keyword argument that specifies the degree of each composite polynomial used in the approximation.
 Orion facilitates PyTorch's ease of use while achieving state-of-the-art FHE neural network performance. While we chose PyTorch, it is straightforward to add support for other modern deep learning libraries.

\vspace{2px}

\noindent\textbf{Range estimation:} Both high-precision bootstrapping and high-degree (Chebyshev) polynomial activation functions require the inputs to be in the range $[-1, 1]$. Thus, when
building large FHE programs, practitioners must insert specific scale-down $\mathsf{PMults}$ to ensure this property holds for all calls to bootstrapping and Chebyshev evaluations. This step is especially necessary for neural networks, which generate intermediate activation values outside of $[-1, 1]$. 

Prior work chooses \textit{per network} scaling factors by manually inspecting and tracking the largest intermediate values seen~\cite{lee2022, Kim2023}. Orion handles this process automatically through \texttt{net.fit()}, which accepts the entire training dataset as input, calculates \textit{per layer} scaling factors, and inserts scale-down multiplications directly into the computational graph. Then, all standard (e.g., ReLU) element-wise activation functions are fit with Chebyshev polynomials, either through interpolation or by the Remez algorithm \cite{minimax}. Listing 1 shows Orion's support for both ReLU and SiLU where the degree(s) of their approximation are passed in as keyword arguments. Extending support to other activation functions is straightforward and follows a process similar to defining custom PyTorch modules. The user need only extend the base \texttt{orion.nn} module, inheriting support for range estimation and polynomial evaluation, and provide an activation function to approximate with a specified degree. In this way, Orion is able to support arbitrary activation functions that can be fit with high-degree polynomials.

\begin{figure}
    \centering
    \includegraphics[width=\linewidth]{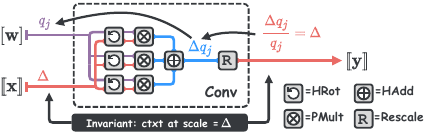}
    \vspace{-15px}
    \caption{An example of Orion’s scale management policy. By encoding plaintexts ($[\mathbf{w}]$) using the appropriate $q_j$ that will be used during rescaling, we are able to maintain the invariant that the scale of any ciphertext must be precisely $\Delta$ between network layers. Each edge color in the graph corresponds to a CKKS polynomial at a specific scale value.}
    \label{fig:scale_management}
    \vspace{-8px}
\end{figure}

\vspace{2px}

\noindent\textbf{Scale management:} Recall from Section \ref{subsubsec:enc} that encoding a cleartext into a CKKS plaintext involves multiplication by a scaling factor, $\Delta$. Properly managing this scale factor for large FHE programs is challenging. To highlight this issue, consider performing a homomorphic multiplication (either $\mathsf{PMult}$ or $\mathsf{HMult}$) where both operands are at level $\ell$ and have a scaling factor of $\Delta$. The resulting ciphertext has a scaling factor of $\Delta^2$ that is then rescaled to $\Delta^2 / q_{\ell} \approx \Delta$ where $q_{\ell}$ is the last prime limb. Thus, rescaling also introduces approximation errors into the resulting ciphertext which grows with circuit depth \cite{errorreduced}. 

In Orion, we propose a new technique to automatically handle scale management that we call \textit{errorless neural network evaluation}. At a high level, we maintain the invariant that the scaling factor of any ciphertext between network layers must be precisely $\Delta$. This technique extends the errorless, depth-optimal polynomial evaluation from Bossuat et al. \cite{DHBSGS} to full neural network inferences. Figure \ref{fig:scale_management} shows an example of our technique: assume we are performing a convolution in Orion. We leverage the fact that our compilation phase has already pre-determined the level at which to perform this convolution (we will call this level $j$). We choose to then encode the kernel weights with scale factor $q_{j}$ (rather than $\Delta$), the last RNS modulus at level $j$. Performing this convolution with a ciphertext at scale $\Delta$ results in an output with scale $\Delta \cdot q_{j}$. Rescaling will divide the ciphertext by the last prime modulus, $q_{j}$, thus precisely resetting this output ciphertext scale back to $\Delta$.

\vspace{2px}

\noindent\textbf{Handling large data structures:} Large datasets and networks require hundreds of gigabytes of rotation keys and matrix diagonals. Orion provides support to store these large data structures to disk in HDF5 format~\cite{hdf5}. Then, the correct data can be loaded dynamically during inference to minimize the size of our transient data.
\vspace{-7px}

\section{Methodology}
\label{sect:methodology}

\noindent{\textbf{CPU setup:}} We evaluate Orion on a \texttt{C4} GCP instance with an Intel Xeon Platinum $8581$C processor clocked at $2.3$ GHz and $512$ GB of RAM. We target Lattigo v$5.0.2$ \cite{lattigo} in our evaluation, an open-source, single-threaded FHE library that supports both double-hoisted matrix-vector products and errorless, depth-optimal polynomial evaluation. Future work includes supporting multi-threaded FHE backends (e.g., OpenFHE \cite{OpenFHE}) and GPU implementations (e.g., HEaaN \cite{HEAAN}). For a fair comparison, we rerun Lee et al. \cite{lee2022} and Fhelipe \cite{fhelipe} on the exact same C4 GCP instance, which are also single-threaded. 

\vspace{2px}

\noindent{\textbf{Activation functions:}} As shown in Table \ref{tab:results}, we use the $x^2$ activation function for MNIST networks and either ReLU or its smoother variant, SiLU \cite{silu}, for all other datasets and networks. We approximate ReLU via a minimax composite polynomial \cite{minimax}. Here, a high degree polynomial is decomposed into a composition of several lower degree polynomials to reduce the number of homomorphic multiplications. Following prior work, our composition consists of three polynomials of degree $15$, $15$, and $27$. For SiLU, we use a degree-$127$ Chebyshev polynomial, obtained using a similar minimax approach. The multiplicative depth of ReLU is $14$ ($13$ for $\mathsf{sign}$($x$)  and $1$ for its multiplication with $x$), whereas the depth of SiLU is just $7$. Later, we will show how the choice of activation function introduces a trade-off in latency and test accuracy. 
\vspace{2px}

\noindent{\textbf{Benchmarks:}} Our evaluation consists of four datasets (image sizes): MNIST ($28 \times 28 \times 1$) \cite{lecun-mnisthandwrittendigit-2010}, CIFAR-10 ($32 \times 32 \times 3$) \cite{krizhevsky2009learning}, Tiny ImageNet ($64 \times 64 \times 3$) \cite{Le2015TinyIV}, and ImageNet ($224 \times 224 \times 3$) \cite{imagenet}. We evaluate three neural networks for MNIST from prior work: a 3-layer MLP from SecureML \cite{secureML}, a 3-layer CNN from LoLA CryptoNets \cite{Brutzkus2019LowLatency}, and the largest LeNet-5 model from CHET \cite{CHET} and EVA \cite{EVA}. For CIFAR-10, we benchmark AlexNet \cite{alexnet}, VGG-16 \cite{vgg}, and ResNet-$20$ \cite{he2015deepresiduallearningimage}, each with ReLU and SiLU activation functions. For Tiny ImageNet, we implement MobileNet-v1 \cite{mobilenet} and ResNet-$18$, and for ImageNet, we evaluate ResNet-$34$ and ResNet-$50$. For all networks, we replace max pooling with average pooling. \label{test:test} Finally, we perform object detection and localization using YOLO-v1 \cite{yolov1} with a ResNet-34 backbone on the PASCAL-VOC dataset ($448 \times 448 \times 3$) \cite{pascalvoc}.

\vspace{2px}

\noindent{\textbf{Validation:}} We validate each benchmark by comparing Orion's FHE outputs against the equivalent cleartext outputs from PyTorch. For MNIST, we perform this validation across all  10,000 test images. For CIFAR-10 (Tiny ImageNet), we instead randomly sample 1,000 ($100$) test images, leading to an estimate of the test accuracy. Finally, for ImageNet, we perform just a single encrypted inference for comparison due to resource constraints.

Alongside accuracy and latency, we also report the mean \textit{precision} (in bits) of the output, which is defined as $- \hspace{-0.1em}\log_2(\epsilon)$, where $\epsilon$ is the mean absolute difference between the outputs of Orion and PyTorch. To compare with prior work, we also report the number of ciphertext rotations and number of bootstrap operations for each network. The links to code blocks for each network and their respective parameter sets can be found in Table \ref{tab:results}.

\vspace{2px}

\setlength{\tabcolsep}{2.70pt}
\setlength{\arrayrulewidth}{0.25mm}
\renewcommand{\arraystretch}{1.3}
\begin{table*}[t]
    \caption{We evaluate Orion on a series of networks and datasets ranging from MLP on MNIST to ResNet-50 on ImageNet. We provide the code, parameter set ($\mathsf{Set}$), rotation amount, number of bootstraps as well as both cleartext and FHE accuracy.}
    \vspace{-7px}
    \small
   \centering
   \begin{tabular}{|c|c|c|c|c|c|c|c|c|c|c|c|c||c|}

    \hline
    \cellcolor{gray!60} & 
    \cellcolor{gray!20}\textbf{Model} & 
    \cellcolor{gray!20}\textbf{Params (M)} & 
    \cellcolor{gray!20}\textbf{FLOPS (M)} & 
    \cellcolor{gray!20}\textbf{Code} & 
    \cellcolor{gray!20}\textbf{Set} & 
    \cellcolor{gray!20}\textbf{\# Rots} & 
    \cellcolor{gray!20}\textbf{Act.} & 
    \cellcolor{gray!20}\textbf{Depth} & 
    \cellcolor{gray!20}\textbf{\# Boots} & 
    \cellcolor{gray!20}\textbf{Clear Acc.} & 
    \cellcolor{gray!20}\textbf{FHE Acc.} & 
    \cellcolor{gray!20}\textbf{Prec. (b)} & 
    \cellcolor{gray!20}\textbf{Time (s)} \\

     \hline \noalign{\vskip 0.1cm}  \hline


    \cellcolor{gray!20}  & MLP & $0.12$  & $0.12$ & \hspace{1px} \large{\href{https://github.com/baahl-nyu/orion/blob/main/models/mlp.py}{\textcolor{black}{\ExternalLink}}} & \hspace{1px} \large{\href{https://github.com/baahl-nyu/orion/blob/main/configs/mlp.yaml}{\textcolor{black}{\ExternalLink}}} & $70$ & $x^2$ & $5$ & $0$ & $98.02\%$ & $98.03\%$ & $4.60$  & $0.29$ \\ \cline{2-14}
    \cellcolor{gray!20}  & LoLA  & $0.10$ & $0.13$ & \hspace{1px} \large{\href{https://github.com/baahl-nyu/orion/blob/main/models/lola.py}{\textcolor{black}{\ExternalLink}}} & \hspace{1px} \large{\href{https://github.com/baahl-nyu/orion/blob/main/configs/lola.yaml}{\textcolor{black}{\ExternalLink}}} & $73$ & $x^2$ & $5$ & $0$ & $98.63\%$ & $98.62\%$ & $4.81$ & $0.23$ \\ \cline{2-14}
    \multirow{-3}{*}{\cellcolor[gray]{0.9}\rotatebox[origin=c]{90}{\textbf{MNIST}}} & LeNet & $1.66$ & $4.30$ & \hspace{1px} \large{\href{https://github.com/baahl-nyu/orion/blob/main/models/lenet.py}{\textcolor{black}{\ExternalLink}}} & \hspace{1px} \large{\href{https://github.com/baahl-nyu/orion/blob/main/configs/lenet.yaml}{\textcolor{black}{\ExternalLink}}} & $282$ & $x^2$ & $7$ & $0$ & $99.31\%$ & $99.31\%$ & $10.4$ & $2.93$ \\

    \hline \noalign{\vskip 0.1cm} \hline

    \cellcolor{gray!20}  & \multirow{2}{*}{AlexNet}    & \multirow{2}{*}{$23.3$} & \multirow{2}{*}{$188$} & \multirow{2}{*}{\hspace{1px} \large{\href{https://github.com/baahl-nyu/orion/blob/main/models/alexnet.py}{\textcolor{black}{\ExternalLink}}}} & \multirow{2}{*}{\hspace{1px} \large{\href{https://github.com/baahl-nyu/orion/blob/main/configs/alexnet.yaml}{\textcolor{black}{\ExternalLink}}}} & \multirow{2}{*}{$1470$} & ReLU & $109$ & $15$ & $92.83\%$ & $92.80\%$ & $4.27$ & $337.2$ \\ \cline{8-14}
    \cellcolor{gray!20}  &                             &                      & & & & & SiLU & $60$ & $7$ & $89.42\%$ & $89.30\%$ & $7.19$ & $190.3$ \\ \cline{2-14}    
    
    \cellcolor{gray!20}  & \multirow{2}{*}{VGG-16}     & \multirow{2}{*}{$14.7$} & \multirow{2}{*}{$314$} & \multirow{2}{*}{\hspace{1px} \large{\href{https://github.com/baahl-nyu/orion/blob/main/models/vgg.py}{\textcolor{black}{\ExternalLink}}}} & \multirow{2}{*}{\hspace{1px} \large{\href{https://github.com/baahl-nyu/orion/blob/main/configs/vgg.yaml}{\textcolor{black}{\ExternalLink}}}} & \multirow{2}{*}{$1771$} & ReLU & $227$ & $28$ & $94.53\%$ & $94.50\%$ & $5.10$ & $588.6$ \\ \cline{8-14}
    \cellcolor{gray!20}  &                             &                      & & & & & SiLU & $137$ & $14$ & $92.27\%$ & $93.60\%$ & $9.72$ & $397.4$ \\ \cline{2-14}  
    
    \cellcolor{gray!20}  & \multirow{2}{*}{ResNet-20}  & \multirow{2}{*}{$0.27$} & \multirow{2}{*}{$41.2$} & \multirow{2}{*}{\hspace{1px} \large{\href{https://github.com/baahl-nyu/orion/blob/main/models/resnet.py}{\textcolor{black}{\ExternalLink}}}} & \multirow{2}{*}{\hspace{1px} \large{\href{https://github.com/baahl-nyu/orion/blob/main/configs/resnet.yaml}{\textcolor{black}{\ExternalLink}}}} & \multirow{2}{*}{$836$} & ReLU & $287$ & $37$ & $93.21\%$ & $93.40\%$ & $4.84$ & $618.2$ \\ \cline{8-14}
    \multirow{-6}{*}{\cellcolor[gray]{0.9}\rotatebox[origin=c]{90}{\textbf{CIFAR-10}}} & & & & & & & SiLU & $154$ & $19$ & $92.61\%$ & $91.70\%$ & $13.6$ & $301.4$ \\ \cline{2-14}  

    \hline \noalign{\vskip 0.1cm} \hline
    

    \cellcolor{gray!20} & MobileNet & $3.25$ & $47.4$ & \hspace{1px} \large{\href{https://github.com/baahl-nyu/orion/blob/main/models/mobilenet.py}{\textcolor{black}{\ExternalLink}}} & \hspace{1px} \large{\href{https://github.com/baahl-nyu/orion/blob/main/configs/mobilenet.yaml}{\textcolor{black}{\ExternalLink}}} & $2508$ & SiLU & $218$ & $42$ & $56.31\%$ & $62.00\%$ & $8.94$ & $892.4$ \\ \cline{2-14}
    \multirow{-2}{*}{\cellcolor[gray]{0.9}\rotatebox[origin=c]{90}{\textbf{Tiny}}} & ResNet-18 & $11.3$ & $2260$ & \hspace{1px} \large{\href{https://github.com/baahl-nyu/orion/blob/main/models/resnet.py}{\textcolor{black}{\ExternalLink}}} & \hspace{1px} \large{\href{https://github.com/baahl-nyu/orion/blob/main/configs/resnet.yaml}{\textcolor{black}{\ExternalLink}}} & $10838$ & SiLU & $138$ & $61$ & $60.57\%$& $57.00\%$ & $8.56$ & $1447$ \\ \cline{2-14}          
                         
    \hline \noalign{\vskip 0.1cm}  \hline
        
    \cellcolor{gray!20} & ResNet-34  & $21.8$ & $3670$ & \hspace{1px} \large{\href{https://github.com/baahl-nyu/orion/blob/main/models/resnet.py}{\textcolor{black}{\ExternalLink}}} & \hspace{1px} \large{\href{https://github.com/baahl-nyu/orion/blob/main/configs/resnet.yaml}{\textcolor{black}{\ExternalLink}}} & $48108$ & SiLU & $267$ & $146$ & $73.66\%$  & N/A & $8.59$& $14338$ \\ \cline{2-14}
    \multirow{-2}{*}{\cellcolor[gray]{0.9}\rotatebox[origin=c]{90}{\textbf{IMNet}}} & ResNet-50  & $25.6$ & $4110$ & \hspace{1px} \large{\href{https://github.com/baahl-nyu/orion/blob/main/models/resnet.py}{\textcolor{black}{\ExternalLink}}} & \hspace{1px} \large{\href{https://github.com/baahl-nyu/orion/blob/main/configs/resnet.yaml}{\textcolor{black}{\ExternalLink}}} & $143217$ & SiLU & $395$ & $351$ & $76.22\%$& N/A & $8.90$ & $32324$ \\         
        
    \hline     
   \end{tabular}
   \vspace{5px}
    \vspace{-15px}
   \label{tab:results}
\end{table*}

\vspace{-8px}

\section{Evaluation}

In this section, we evaluate Orion, quantify our improvements over prior work, and demonstrate the effectiveness of our approach. We begin by presenting our evaluations across all networks and datasets in Table \ref{tab:results} that highlight the benefits of our single-shot multiplexing strategy as well as our automatic bootstrap placement algorithm. Next, we analyze the efficiency of our automatic bootstrap placement algorithm as we increase network depth, and finally we close with an object detection and localization case study. To the best of our knowledge, this is the first high-resolution ($448 \times 448 \times 3$) object detection using a deep neural network in FHE.

\setlength{\tabcolsep}{3.5pt}
\setlength{\arrayrulewidth}{0.25mm}
\renewcommand{\arraystretch}{1.3}
\begin{table}[t]
   \caption{A comparison of ciphertext rotation counts in CIFAR-10 networks between Lee et al. \cite{lee2022} and Orion.}
    \vspace{-7px}
    \small
   \centering
   \begin{tabular}{|c|c|c|c|c|}
   \hline
   \cellcolor{gray!20}\textbf{Work} & 
   \cellcolor{gray!20}\textbf{ResNet-20} & 
   \cellcolor{gray!20}\textbf{ResNet-110} & 
   \cellcolor{gray!20}\textbf{VGG-16} & 
   \cellcolor{gray!20}\textbf{AlexNet} \\

   \hline \noalign{\vskip 0.1cm} \hline

   Lee et al. \cite{lee2022} & $1382$ & $7622$ & $9214$ & $9422$ \\ \hline
   Orion (us) & $836$ & $4676$ & $1771$ & $1470$ \\ \hline
   Improvement & $1.65 \times$ & $1.64 \times$ & $5.20 \times$ & $6.41 \times$ \\ \hline
    
   \end{tabular}
   \label{tab:rotation_counts}
   \vspace{-10px}
\end{table}

\vspace{-8px}

\subsection{MNIST}
\label{subsec:mnist}

For MNIST, we use the conjugate invariant \cite{conjugateinvariant} ring type in CKKS to set the number of slots equal to the ring degree, as opposed to the traditional $n=\sfrac{N}{2}$ used when bootstrapping is required. All other benchmarks (CIFAR-10 and beyond) use the traditional ring type where $n = \sfrac{N}{2}$. Notably, since our single-shot multiplexed convolutions consume only one level, the depth of networks such as LoLA are also roughly halved when compared to prior work. For instance, the LoLA implementation in Fhelipe \cite{fhelipe} (PLDI '24) has a multiplicative depth of $10$, whereas in Orion its depth is just $5$. This enables us to reduce the ring degree from the typical $N=2^{14}$ to $N=2^{13}$ while remaining 128-bit secure. As a result, we improve upon the results of Fhelipe by nearly $83 \times$, reducing end-to-end latencies from $19.0$ seconds to just $0.23$ seconds. With the same parameter set as Fhelipe, we achieve a mean latency of $ 0.97$ seconds ($19 \times$ reduction). Similarly, our LeNet-$5$ latency of $2.93$ seconds is roughly $44 \times$ faster than the single-threaded results from EVA \cite{EVA}.

\vspace{-5px}

\subsection{CIFAR-10}
\label{subsec:cifar}

\noindent\textbf{Packing comparisons:} Table \ref{tab:rotation_counts} more concretely compares our single-shot multiplexing strategy against the multiplexed approach from Lee et al. \cite{lee2022} using the CIFAR-10 networks from Table \ref{tab:results} alongside ResNet-$110$. Notably, our improvement over prior work \textit{increases} with model complexity. This improvement occurs for two reasons. First, the benefits of BSGS increase with filter size since, from Section \ref{sect:packing_convolutions}, the complexity of homomorphic convolutions decreases from $\mathcal{O}(f)$ to $\mathcal{O}(\sqrt{f})$, with $f$ the number of filter elements. Second, for small networks such as ResNet-20, we rely on Gazelle's hybrid method to diagonalize matrices that are often much smaller than $n \times n$, where $n$ is the number of slots. Doing so maintains the property that convolutions only consume one level, however it induces sparser plaintext diagonals. For larger, multi-ciphertext networks that Orion primarily targets, the hybrid method is not needed, and plaintext diagonals are packed as densely as possible. These two reasons are also why AlexNet, despite having $86\times$ the number of parameters as ResNet-$20$, only has $1.76\times$ more rotations.

\vspace{2px}

\noindent\textbf{Performance comparisons:} Table \ref{tab:fhelipe_comparison} further highlights our sources of improvement over the prior work of Fhelipe \cite{fhelipe} when run on the same GCP instance. Notably, despite having only $1.71\times$ fewer ciphertext rotations, our convolutional runtime is $11.2 \times$ faster which occurs for two reasons. First, \textit{all} ciphertext rotations in Orion are performed with double-hoisting. Recall from Section \ref{sect:mv_products} that hoisting amortizes the expensive aspects of the key-switch procedure \textit{across} many ciphertext rotations and is only possible when using the diagonal encoding method. Second, Orion's compilation phase automatically generates and stores all rotation keys and encoded matrix diagonals. On the other hand Fhelipe generates all encoded plaintexts on-the-fly \textit{during} each convolution. The former is a better strategy, even if data must be stored to disk, since CKKS encoding involves both the iFFT and NTT. 

\setlength{\tabcolsep}{4.3pt}
\setlength{\arrayrulewidth}{0.25mm}
\renewcommand{\arraystretch}{1.3}
\begin{table}[t]
   \caption{Quantifying the several sources of improvement in ResNet-20 performance over the prior work of Fhelipe \cite{fhelipe}.}
   \vspace{-7px}
    \small
   \centering
   \begin{tabular}{|c|c|c|c||c|}
   \hline
   \cellcolor{gray!20}\textbf{Work} & 
   \cellcolor{gray!20}\textbf{\# Rots.} & 
   \cellcolor{gray!20}\textbf{\# Boots.} & 
   \cellcolor{gray!20}\textbf{Convs. (s)} & 
   \cellcolor{gray!20}\textbf{Latency (s)} \\

   \hline \noalign{\vskip 0.1cm} \hline

   Fhelipe \cite{fhelipe} & $1428$ & $58$ & $334.5$ & $1468$ \\ \hline
   Orion (us) & $836$ & $37$ & $29.89$ & $618.2$ \\ \hline
   Improvement & $1.71 \times$ & $1.58 \times$ & $11.2 \times$ & $2.38 \times$ \\ \hline
    
   \end{tabular}
   \label{tab:fhelipe_comparison}
   \vspace{-10px}
   \vspace{-5px}
\end{table}

\vspace{2px}

\noindent\textbf{Choice of activation function:} Finally, we explore the trade-off in latency and accuracy by using different activation functions. In more detail, SiLU consumes half the levels of ReLU, which reduces the total multiplicative depth of the circuit. And in turn, less bootstraps are required during inference. We find that using SiLU decreases the cleartext accuracy on average by $2.1\%$ but results in a $1.77 \times$ average speedup. This trade-off is straightforward to further explore given Orion's native support for low-degree polynomial activation functions with \texttt{on.Activation()}. For larger experiments (e.g., Tiny ImageNet and ImageNet), we opt to train our models with SiLU to reduce multiplicative depth and decrease FHE runtime.

\vspace{-7px}

\subsection{Tiny ImageNet}

We now present the results of our Tiny ImageNet experiments on MobileNet-v1 and ResNet-18. Prior work does not run inference using Tiny ImageNet; we report our runs in Table \ref{tab:results}. Despite both networks having fewer parameters than VGG-16 and AlexNet, the number of ciphertext rotations increases substantially. This occurs because the number of ciphertext rotations is more closely tied to the network's FLOPS (number of floating point operations), than its parameter count. Since the input image size has increased four-fold from $(32 \times 32 \times 3)$ to $(64 \times 64 \times 3)$, the size of our matrix-vector products has also increased by the same amount.

While MobileNet is a \textit{deeper} network than ResNet-18, our automatic bootstrap placement algorithm places \textit{fewer} total bootstraps in it. This occurs because MobileNet has no residual connections, while ResNet-18 has eight. Residual connections place an additional constraint on our bootstrap placement algorithm, typically that the input and output levels of the residual block be the same and doing so generally increases bootstrap counts. This same observation is found in Baruch et al. \cite{helayers2}. Furthermore, since MobileNets are designed for mobile and embedded systems, they contain much cheaper depth-wise separable convolutions. Looking closer, we find that the average level that Orion performs MobileNet convolutions at is $\ell=6$, whereas in ResNet-18 it is $\ell=3$. This indicates that our bootstrap algorithm is more aggressively trading off the runtime of (cheaper) convolutions for fewer bootstraps in MobileNet than it is in ResNet-18. 

\vspace{-8px}

\subsection{ImageNet}
To demonstrate both the scalability and user-friendliness of Orion, we evaluate ResNet-34 and ResNet-50 on the ImageNet-1k dataset. Importantly, Orion does not require any FHE-specific training or modification to these networks such as the removal of skip connections or range-aware loss functions (both of which are explored in HeLayers \cite{helayers2}). Additionally, these networks have $81 \times$ and $95 \times$ more parameters than the largest networks supported by Fhelipe \cite{fhelipe}. Here, we directly load the pretrained weights from \texttt{torchvision} and simply finetune both networks after replacing ReLU activations with SiLU and max pooling with average pooling. Since Orion extends PyTorch, this fine-tuning can be performed using existing PyTorch training scripts, and our SiLU models match the accuracy of \texttt{torchvision}'s ReLU models.

The (\textit{single-threaded}) end-to-end encrypted inference time for both networks are $3.98$ hours and $8.98$ hours, respectively. Baruch et al. \cite{helayers2} evaluate a similar ResNet-50 model across $32$ CPU threads using HEaaN \cite{HEAAN} in $2.53$ hours by replacing ReLU with its degree-$18$ approximation. Since a direct comparison is challenging, we note that our ResNet-50 implementation contains just $351$ bootstrap operations, whereas they use 8,480 bootstraps \cite{helayers2}. While we could only evaluate one encrypted inference per network, our results match the cleartext PyTorch output with $8$ bits of precision.

\vspace{-5px}

\subsection{Bootstrap Placement Complexity}
Our automatic bootstrap placement algorithm scales well with both network depth and complexity.  Table \ref{tab:boot_scalability} presents the compile time, bootstrap placement time, and number of bootstraps for ResNet-20 through ResNet-110 using the same composite approximation to ReLU from Section \ref{sect:methodology}. Bootstrap placement time refers to the time our algorithm takes to determine the location of every bootstrap in the network. 

We find a linear increase in bootstrap placement time as network depth increases. We explain this by extending the fully-connected network in Figure \ref{fig:auto_bootstrap}b to an arbitrary depth $d$. Each layer contains  $L_\text{eff} + 1$ vertices ($V$) and is connected to the next layer through $L_\text{eff}^2$ edges ($E$), where $L_\text{eff}$ is the number of levels remaining after each bootstrap operation. The shortest path, found using topological sorting and relaxation, has complexity $\mathcal{O}(|V|+|E|)$ and therefore grows linearly with network depth as $\mathcal{O}(L_\text{eff}^2 \cdot d)$. 

Our bootstrap placement algorithm remains tractable in residual networks such as ResNets since skip connections are always constrained within a residual block and are non-overlapping. Thus, their single-entry, single-exit (SESE) regions can be black-boxed and solved independently. As an example, a solution to ResNet-32 requires black-boxing only six additional SESE regions compared to ResNet-20, with one extra region per residual block. Performance can be further improved by parallelizing this black-boxing stage.

\vspace{-5px}

\setlength{\tabcolsep}{3.4pt}
\setlength{\arrayrulewidth}{0.25mm}
\renewcommand{\arraystretch}{1.3}
\begin{table}[t]
   \caption{An analysis of the scalability of our automatic bootstrap placement algorithm with network depth on ResNets.}
   \vspace{-5px}
    \small
   \centering
   \begin{tabular}{|c|c|c|c|c|c|}
   \hline
   \cellcolor{gray!20}\textbf{Operation} & 
   \cellcolor{gray!20}\textbf{Res-20} & 
   \cellcolor{gray!20}\textbf{Res-32} & 
   \cellcolor{gray!20}\textbf{Res-44} & 
   \cellcolor{gray!20}\textbf{Res-56} & 
   \cellcolor{gray!20}\textbf{Res-110} \\

   \hline \noalign{\vskip 0.1cm} \hline

   Compile (s) & $437$ & $654$ & $867$ & $1096$ & $2132$ \\ \hline
   Boot. Place. (s) & $1.94$ & $2.91$ & $3.86$ & $5.70$ & $11.0$ \\ \hline
   \# Bootstraps & $37$ & $61$ & $85$ & $109$ & $217$ \\ \hline
    
   \end{tabular}
   \vspace{-10px}
   \label{tab:boot_scalability}
\end{table}

\begin{figure}
    \centering
    \includegraphics[width=1\linewidth]{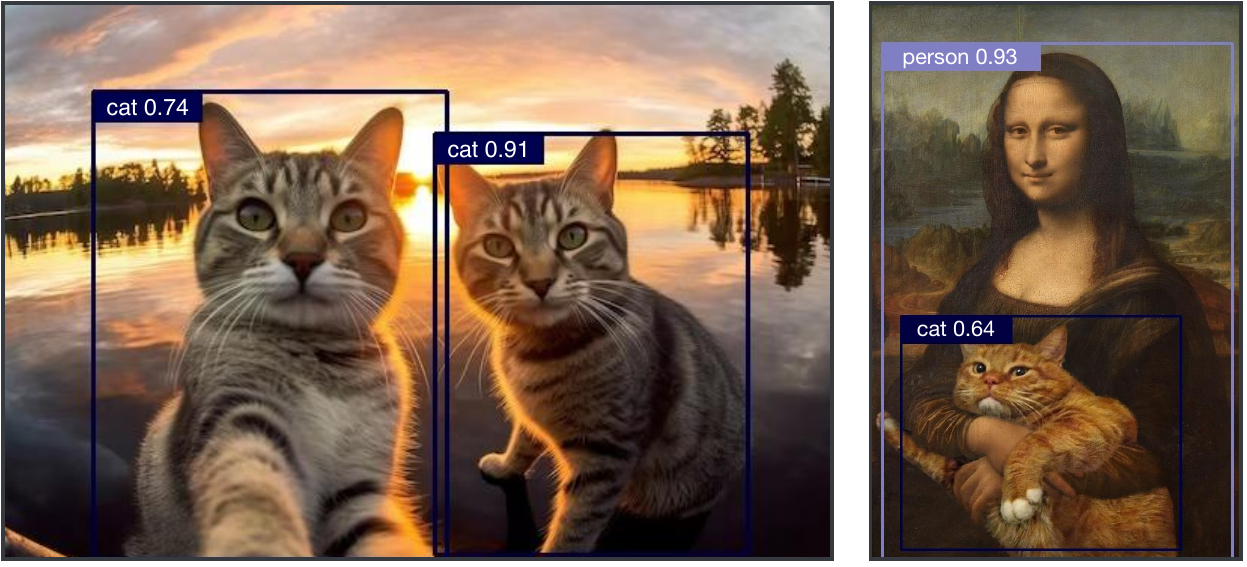}
    \vspace{-18px}
    \caption{The first homomorphic object detection and localization results. Labels indicate each object's predicted class and its confidence score in $[0,1]$.}
    \label{fig:object_detection}
     \vspace{-13px}
\end{figure}

\subsection{Case Study: Object Localization}

We close our evaluation with the first large-scale homomorphic object detection and localization experiments. In these experiments, we train a YOLO-v1 \cite{yolov1} model with a ResNet-34 backbone on the PASCAL-VOC dataset \cite{pascalvoc}, which consists of $20,\hspace{-0.1em}000$ images, each resized to $448 \times 448 \times 3$, spanning $20$ classes. Our model has $139$ million parameters and is designed to predict instances of these $20$ classes within an image along with their corresponding bounding boxes.

Figure \ref{fig:object_detection} visualizes two FHE output predictions from Orion, each with single-threaded latencies of $17.5$ hours. While this is still far from practical, its implementation took just $60$ additional lines of code (\hspace{0.1em}{\Large{\href{https://github.com/baahl-nyu/orion/blob/main/models/yolo.py}{\textcolor{black}{\ExternalLink}}\hspace{-0.4em}}} ) and required minimal changes from its analogous PyTorch implementation.

\vspace{2px}

\vspace{-6px}
\section{Related Works}

Prior work in FHE programming falls into two major categories: circuit-level compilers and domain-specific compilers. Circuit-level compilers \cite{ramparts, alchemy, heaanmilr, hecate, elasm, e3, viand2023heco, coyote, porcupine, progsynth, google-transpiler, marble, chocotaco} represent most of prior work and focus on lower-level optimizations for general programs and typically target smaller workloads. In contrast, domain-specific compilers (e.g., for machine learning) \cite{CHET, EVA, dacapo, fhelipe, ngraph-he, ngraph-he2, tenseal, sealion} often sacrifice this fine-grained control to efficiently support much larger applications. Orion falls into the latter category.

\vspace{2px}

\noindent\textbf{Circuit-level compilers:} Early FHE compilers \cite{alchemy, armadillo, e3} focused on circuit-level optimizations, and more recent tools such as Porcupine \cite{porcupine}, Coyote \cite{coyote}, and HECO \cite{viand2023heco}, offer automated solutions for scheduling FHE instructions. HECO adopts the MLIR framework to target a wide variety of FHE backends and hardware. In parallel, Hecate \cite{hecate} and ELASM \cite{elasm} propose several rescaling techniques to both improve performance and explore the tradeoff in scale management and latency. However, as the authors note, scaling these techniques to deep learning workloads remains infeasible.

\vspace{2px}

\noindent\textbf{Domain-specific compilers:} CHET \cite{CHET} was one of the first FHE compilers to target machine learning workloads. Its focus included automatically selecting encryption parameters, data layouts, and introduced an intermediate representation to decouple program execution from improvements to cryptography. EVA \cite{EVA} improved upon CHET by proposing the waterline rescaling technique to efficiently manage scaling factors. Both CHET and EVA were implemented in SEAL \cite{sealcrypto}, which does not natively support bootstrapping and therefore they do not target \textit{deep} neural networks. nGraph-HE \cite{ngraph-he, ngraph-he2}, TenSEAL \cite{tenseal}, and SEALion \cite{sealcrypto} also target machine learning with Python APIs, but similarly lack the the ability to automate bootstrapping.

Recently, Dacapo \cite{dacapo} proposed an automatic bootstrap placement algorithm and integrated their solution into the GPU-accelerated HEaaN library \cite{ckks}. Their approach involves computing a set of candidate bootstrap locations and then estimating the latency for bootstrapping at different combinations of these locations. HeLayers \cite{helayers1} also automates bootstrap placement and further provides a robust framework for deep learning inference in the most popular FHE backends \cite{lattigo, OpenFHE, ckks}. To the best of our knowledge, this is the only FHE compiler outside of Orion to support datasets larger than CIFAR-10, and it similarly supports ImageNet.

Fhelipe \cite{fhelipe} is the closest prior work to Orion. Although a more general compiler for tensor arithmetic, it is also capable of supporting deep learning applications. Unlike Fhelipe, the goal of Orion is not to compile FHE programs into a list of primitive operations. Instead, we target a higher level of abstraction (e.g., linear transforms) to leverage cryptographic optimizations such as hoisting in our FHE backend. It is here that we find the majority of our performance improvements. 

\vspace{-5px}
\vspace{-8px}

\section{Conclusion}
In this paper, we present \textit{Orion}, a framework that completely automates and translates neural networks directly into FHE programs. Orion allows both researchers and practitioners to rapidly iterate on their ideas and understand FHE using a high-level machine learning library such as PyTorch. We propose  our \textit{single-shot multiplexed} packing strategy that implements arbitrary convolutions and  our automatic bootstrap placement algorithm that requires no user input. We integrate both directly into Orion and achieve state-of-the-art (single-threaded) latency for standard FHE benchmarks. Orion can run large-scale neural networks such as ResNet-50 on ImageNet and even YOLO-v1 object detection on images of size $448 \times 448 \times 3$. 

Going forward, we plan to lower Orion to alternative backends such as multi-threaded FHE libraries (e.g., OpenFHE \cite{OpenFHE}) and GPU systems (e.g., HEaaN-GPU \cite{ckks}, Cheddar \cite{cheddar}, ArctyrEX \cite{arctyrex}). Additionally, our high-level Python interface allows other researchers to extend Orion to support new networks layer types such as self-attention. By lowering the barrier to entry into this field, Orion helps strengthen and embolden research within the FHE community. 

\vspace{-8px}

\section*{Acknowledgements}
This work was supported in part by Graduate Assistance in Areas of National Need (GAANN). The research was developed with funding from the NSF CAREER award \#2340137 and DARPA, under the Data Protection in Virtual Environments (DPRIVE) program, contract HR0011-21-9-0003. Reagen and Ebel received generous support from the NY State Center for Advanced Technology in Telecommunications (CATT) and a gift award from Google. We especially thank Jean-Philippe Bossuat and the anonymous reviewers for their thoughtful feedback. The views, opinions, and/or findings expressed are those of the authors and do not necessarily reflect the views of sponsors.

\label{sec:references}


\end{document}